\documentstyle[draft,prl,aps,twocolumn,epsfig]{revtex}
\newcommand{\be}{\begin{equation}}
\newcommand{\ee}{\end{equation}}
\newcommand{\r}{\frac{r}{R}}
\newcommand{\lp}{\left(}
\newcommand{\rp}{\right)}
\newcommand{\la}{\left\langle}
\newcommand{\ra}{\right\rangle}

\begin{document}
\bibliographystyle{prsty}

\title{Multiscale velocity correlations in turbulence }
\date{\today}
\author{R. Benzi$^{1}$, L. Biferale$^{2,4}$
   and F. Toschi$^{3,4}$ }
\address{$^{1}$ AIPA, Via Po 14, 00100 Roma, Italy.\\
         $^{2}$  Dipartimento di Fisica, Universit\`{a} di Tor Vergata,
         Via della Ricerca Scientifica 1, I-00133 Roma, Italy.\\
         $^{3}$ Dipartimento di Fisica, Universit\`a di Pisa,
           Piazza Torricelli 2, I-56126 Pisa, Italy.\\
         $^{4}$  INFM-Unit\'a di Tor Vergata.}
\maketitle

\begin{abstract}
Multiscale correlation functions in high Reynolds number experimental
turbulence and synthetic signals are investigated. \\
Fusion Rules predictions as they arise from multiplicative,
almost uncorrelated, random processes for the energy cascade are tested.\\
Leading and sub-leading contribution, in both the inertial
and viscous ranges,  are well captured 
by assuming a simple multiplicative random process for
the energy transfer mechanisms.\\
\end{abstract}

In stationary turbulent flows, a net flux of energy establishes in the 
inertial range, i.e. from forced scales, $L$,  down to the
dissipative scale $r_d$. 
Energy is transferred through a statistically scaling-invariant process,
which is characterized by a strongly non-gaussian (intermittent)
activity.
Understanding the statistical properties of intermittency is
one of the  most  challenging open problem in three dimensional fully
developed turbulence.\\

Intermittency in the inertial range  is usually analyzed by means of 
the statistical
properties of  velocity differences, $\delta_r v(x) =v(x) -v(x+r)$. 
In particular, in the last twenty years \cite{frisch}, overwhelming
experimental and theoretical works focused on structure functions:
$S_p(r) =  \la (\delta_r v(x))^p \ra$. 
A wide agreement exists on the fact
that structure functions show a scaling behaviour in the limit of
very high Reynolds numbers, i.e. in presence of  a large
separation between integral and dissipative scales, 
$L/r_d \rightarrow \infty$:
\be
S_p(r) \sim \lp {r \over L} \rp^{\zeta(p)}
\label{structure}
\ee
The velocity fluctuations are anomalous in the sense that 
$\zeta(p)$ exponents do not follows the celebrated dimensional
prediction
made by Kolmogorov, $\zeta(p)=p/3$. In fact, 
 $\zeta(p)$ are observed  to be a nonlinear
function of $p$, which is the most important signature of the 
intermittent transfer of fluctuations from large to small scales.\\
In order to better characterize the transfer mechanism, it is natural to
look also at correlations among velocity fluctuations at different
scales. 
Multiscale correlations functions should play in turbulence the same 
r\^ole played by correlation functions in critical statistical
phenomena. 

\noindent
Recently,  some theoretical work \cite{pro1,pro2} and an 
exploratory experimental investigation \cite{sreene} have been devoted 
to the behavior of multiscale velocity correlations:
\begin{eqnarray}
F_{p,q}(r,R) &\equiv&  \langle \lp v \lp x+r 
\rp-v \lp x\rp \rp^p \lp v\lp x+R\rp-v\lp x\rp\rp^q  \rangle  \nonumber \\
&\equiv& \langle (\delta_r v(x))^p(\delta_R v(x))^q \rangle  
\label{2s}
\end{eqnarray}
with $r_d <  r  <  R  < L$.  When 
the smallest among the two scales $r$ goes
 beyond  the dissipative scales, $r_d$, new properties of the
correlation
functions (\ref{2s}) may arise due to the non trivial physics of the
dissipative cutoff. From
now on, we will mostly concentrate on correlation functions with both
$r$ and $R$
in the inertial range. Moreover, in order to simplify our discussion, we
will confine our analysis 
for the case of longitudinal velocity differences.\\

Stochastic cascade processes  are simple and 
well known useful  tools to describe the leading phenomenology of the
intermittent 
energy transfer in the inertial range. Both anomalous scaling 
exponents and viscous effects \cite{frisch,physicad96} can be 
reproduced by choosing a suitable
random process for the multiplier, $W(r,R)$,  which connects 
velocity fluctuations  at two different scales, $R >  r$. \\
The main finding of this letter is that experimental multiscale
correlations (\ref{2s})  are in {\it quantitative} agreement,
for any separation of scale $r/R$,  with the prediction
one obtains by using a pure uncorrelated multiplicative
process for the energy cascade. 

The main idea turns around the hypothesis that small scale statistics is
fully determined by a cascade process conditioned  to  some
large scale configuration:
\be
\delta_rv(x) = W(r,R) \cdot \delta_Rv(x)
\ee
where, requiring homogeneity along the cascade process, the
random function $W$ should depend only on the ratio $r/R$.
Structure functions are then described in terms of the $W$ process:
$S_p(r) = C_p  \langle  \left[ W\left( r/L \right)\right]^p  \rangle $, where 
$C_p =  \langle  (\delta_{L}v(x))^p  \rangle $ 
if the stochastic multiplier may be
 considered almost uncorrelated with the large-scale velocity field. 
Pure power laws arise in the high Reynolds regime: in this limit we must
 have  $ \langle  [W(\r)]^p  \rangle  \sim (\r)^{\zeta(p)}$. 
In the same framework, it is straightforward
 to give the leading prediction for 
the multiscale correlation functions (\ref{2s}):
\begin{equation} 
F_{p,q}(r,R) \sim  \la 
 \left[ W\left( \rule[-2ex]{0cm}{4ex} \r \right) \right]^p 
\left[W \left(\frac{R}{L} \right)\right]^{p+q}  \ra,
\label{frf}
\end{equation}
which becomes in the hypothesis of negligible correlations among
multipliers:
\begin{eqnarray}
F_{p,q}(r,R) &=& C_{p,q}  \la  \left[W\left( \rule[-2ex]{0cm}{4ex} 
\r\right)\right]^p  \ra   
\la  \left[W\left(\frac{R}{L}\right)\right]^{p+q}  \ra  \sim\nonumber \\
&\sim& \frac{S_p(r)}{S_p(R)} \cdot S_{p+q}(R)
\label{fritamar}
\end{eqnarray}

This expression was for the first time proposed in \cite{pro1} and 
considered to rigorously express the leading behavior of
 (\ref{2s}) when
$r/R \rightarrow 0$ as long as some weak hypothesis of scaling 
invariance and of universality of scaling exponents  
in Navier-Stokes equations  hold.
Let us notice that, beside any rigorous claim,
expression (\ref{fritamar}) is also the zero-{\it th} order prediction
starting
from any multiplicative uncorrelated random cascade satisfying 
$ \langle  [W(\r)]^p  \rangle  \equiv  S_p(r)/S_p(R)$. \\

In this letter we want to address three main questions: (i) whether the 
prediction (\ref{fritamar}) gives the correct leading behavior in  the 
limit of large separation of scales $r/R \sim 0$, 
(ii) if this is the case, what one can 
say about sub-leading behavior for  separation 
$r/R \sim O(1)$, (iii) what happens to those observables  for which the
"multiplicative prediction" (\ref{fritamar}) is incorrect 
because of symmetry reasons. Indeed, let 
us notice that  for correlation like :
\be
F_{1,q}(r,R)=  \la \left(\delta_{r}v \right) \left(\delta_{R}v\right)^{q} \ra 
\label{fr1}
\ee
the multiplicative prediction gives: 
$$ F_{1,q}(r,R)= {{S_1(r)} \over {S_1(R)}} \cdot 
S_{1+q}(R).$$ Such a prediction is 
wrong because, if homogeneity can be assumed, $S_1(r)=0$ for all scales 
$r$.  In this case prediction (\ref{fritamar})
does not represent the leading contribution.

In this letter we propose a systematic investigation of (\ref{2s}) in 
high Reynolds number experiments and synthetic 
signals. 
The main purpose consists in  probing whether
multiscale correlation functions may show new dynamical
properties (if any) which are not taken into account by the
standard simple multiplicative models for the energy transfer.

Experimental data have been obtained in a wind tunnel (Modane)
with $Re_{\lambda}=2000$. The integral scale is
$L\sim  20 \,m$ and the dissipative scale is $r_d = 0.31\, mm$.
Synthetic signals are built in terms of
wavelet decomposition with coefficients defined by
a pure uncorrelated random multiplicative process \cite{bbcpvv}.
Such a  signal should therefore show the strong fusion rules 
prediction (\ref{fritamar}) and it will turn out to be an useful
tool for testing how much deviations from (\ref{fritamar}), observed
in experiments or numerical simulations, are due to important
dynamical effects or only to unavoidable  geometrical corrections.

First of all, let us notice that for any 1-dimensional string of number
(such as the
typical outcome of laboratory experiments in turbulence) the multiscale
correlations (\ref{2s}) feel strong geometrical 
constraints. In particular we may
always write down  "Ward-Identities" (WI):
\begin{eqnarray}
S_p(R-r) &\equiv&  \la \left[(v(x+R)-v(x))-(v(x+r)-v(x))\right]^p
 \ra  \noindent \\
&=&
\sum_{k=0,p}b(k,p)(-)^k F_{k,p-k}(r,R),
\end{eqnarray}
where $b(k,p)=p!/[k!(p-k)!]$.\\
For example, for $p=2$ we have 
\begin{eqnarray}
2 F_{1,1}(r,R) &\equiv& S_2(r) + S_2(R) -S_2(R-r) \sim\nonumber \\
& \sim& \left[\left(\r\right)^{\zeta(2)}+ O\left(\r\right)\right] \cdot S_2(R)
\label{f11}
\end{eqnarray}
where the latter expression has been obtained by expanding $S_2(R-r)$
in the limit $r/R \rightarrow 0$.\\
For $p=3$ we have 
$$S_3(R-r) = S_3(R) -S_3(r) +3 F_{2,1}(r,R) -3F_{1,2}(r,R)$$

The "Ward-Identities" will turn out to be useful for understanding
sub-leading predictions
to the multiplicative cascade process. One may argue that 
in geometrical set-up different from the one specified in (\ref{2s})
the same kind of
constraint will appear with eventually different weights among 
different terms.

The main result presented in this letter is that all multiscale
correlations 
functions 
are well reproduced in their leading term, $\r \rightarrow 0$,
by a simple uncorrelated random 
cascade (\ref{fritamar}) and that   their sub-leading contribution,  
$ \r \sim O(1)$, are fully captured
by the geometrical constrained 
previously discussed, namely the "Ward-Identities".
 
The recipe for calculating multiscale correlations will  
be the following:
first, apply the multiplicative guess for the leading contribution and
look
for geometrical constraints in order to find out sub-leading terms.
Second, in all cases where the leading multiplicative contribution
vanishes because of underlying symmetries,
look directly for the geometrical constraints and find out what is
the leading contribution
applying the multiplicative random approximation to all, non-vanishing,
terms
in the  WI.

Let us  check the strong fusion rules prediction (\ref{fritamar}) for
moments with $p>1,q>1$. 
In Figure 1 we have checked the large scale dependency  by plotting
$ F_{p,q}(r,R)/S_{p+q}(R) \cdot S_p(R)$ as a function of $R$ at 
fixed, $r$, for different values of $p,q$. \\
The expression (\ref{fritamar}) predicts  the existence of a plateau
(independent of $R$)
at all scales $R$ 
where the leading multiplicative description is correct. 

>From Figure 1 one can see that, in the limit of large separation $R
\rightarrow L$ at fixed $r$,
$ F_{p,q}(r,R)/S_{p+q}(R) \cdot S_p(R)$ shows a tendency toward a plateau. On
the other hand, there are clear deviations 
for $r/R \sim O(1)$. Such deviations show a very
slow decay as a function of the scale separation.\\

In order to understand the physical meaning of the observed deviations
to the fusion rules (\ref{fritamar}),
we compare, in Figure 1, the experimental data against the equivalent
quantities measured by using 
the synthetic signal.
We notice an almost perfect superposition
of the two data sets,
indicating that the deviations observed in real data 
can hardly be considered a "dynamical effect".\\

Using the WI plus our multiplicative receipt
for $p=4$ we quickly read that the leading contribution
to $F_{2,2}$ is $ O(r^{\zeta(2)}) \cdot O(R^{\zeta(4)-\zeta(2)})$, while
sub-leading terms scale as   $O(r^{\zeta(4)})$, and as 
$O(r^{\zeta(3)}) \cdot O(R^{\zeta(4)-\zeta(3)})$.\\
This superposition of power laws 
is  responsible for the slowly-decaying correlations in Fig. 1.
The result so far obtained, i.e.  that both the experimental data and the
synthetic signal show the same quantitative behaviour, is a strong indication
that  multiscale correlation functions, at least for  $p > 1, q> 1$,
 are in good agreement
with the random multiplicative model for the energy transfer. \\

For multiscale correlations where the direct
application of the random-cascade
prediction is useless,  like $F_{1,q}(r,R)$, we use 
the WI plus the multiplicative prediction applied to all terms, except 
the $F_{1,q}$. One obtains the  expansion:
\begin{eqnarray}
F_{1,q}(r,R) &\sim& \left[ O\left( \r \right)^{\zeta(2)} + 
O\left(\r \right)^{\zeta(3)}
+O\left(\r\right)^{\zeta(4)} + \nonumber \right. \\ 
 &\cdots& \left.+ \; O\left(\r\right)^{\zeta(q+1)} \right] \cdot S_{q+1}(R),
\label{f1q}
\end{eqnarray}
which coincides when $q=1$ with the exact result (\ref{f11}) using 
$\zeta(3)=1$.

In Fig. 2 we show the experimentally measured $F_{1,2}$ and the fit that
we obtain by keeping only the first two terms of the 
expansion in (\ref{f1q}). The fit has been performed by  imposing the
value for the scaling exponents $\zeta(2), \, \zeta(3)$ measured
on the structure functions, i.e. only the coefficients
in front of the power laws have been fitted. 
As one can notice, the fit
works perfectly in the inertial range.
Let us remark that 
the correlation changes sign in the middle 
of the inertial range, which 
is a clear indication that a single power-law
fit (neglecting sub-leading terms) would completely miss the correct
behaviour. 

Next we consider the WI for $p=3$. Due to the fact that $S_3(r) \sim r$
in the inertial range,
one can easily show that the WI
forces $F_{12} \sim F_{21}$. Therefore we can safely state that 
also correlation functions of the form $F_{p,1}$
feel non trivial dependency from the large scale $R$,
at variance with prediction given in \cite{pro2}
using isotropic arguments.

Let us summarize what is the framework we have found until now. \\
Whenever the simple scaling ansatz based on the 
uncorrelated multiplicative process is not prevented by
symmetry arguments, the multi-scale correlations are in good
asymptotic agreement with the fusion rules prediction even if
strong corrections due to sub-leading terms are
seen for small-scale separation $r/R \sim O(1)$.
Subleading terms are strongly connected to the WI  previously
discussed, i.e. to geometrical constraints.
In the other cases (i.e. $F_{1,q}(r,R)$)
the geometry fully determines both leading and  sub-leading scaling.\\
All this findings, led us to the conclusions that multiscale
correlations 
functions measured in turbulence are fully consistent
with a multiplicative, almost uncorrelated, process.\\
Nevertheless, the strong and slowly-decaying sub-leading corrections
to  the naive multiplicative
fusion rules predictions are particularly annoying for any attempts  to
attack 
analytically  the
equation of  motion for structure functions; in that case, multiscale
correlations at almost coinciding scales 
are certainly the dominant contributions 
in the non-linear part of the equations \cite{pro2}. Indeed, as
shown in an  analytical calculation 
for a dynamical toy model of random passive-scalar advection \cite{bbw},
fusion rules are violated at small scale-separation and 
the violations are relevant for correctly evaluating the exact behavior
of structure functions at all scales.

When the smallest distance $r$ is inside the viscous length,
one can use the approach of multiplicative processes with multiscaling
viscous cutoff \cite{fv}. Namely, for the correlation 
$D_{1,q}(R) =  \langle  (\partial_xv)^{2} (\delta_R v)^q  \rangle$ one
obtains:
\begin{equation}
D_{1,q}(R) \sim  \la  \left(\delta_R v \right)^q  
\left(\frac{\delta v(r_d)}{r_d}\right)^2 \ra 
\end{equation}
where $r_d$ is the dissipative scale.
In the multifractal interpretation we say: 
$\delta_{r_d} v = (r_d/R)^h \cdot \delta_R v$
with probability $P_h(r_d,R)= (r_d/R)^{3-D(h)}$.
 Following \cite{fv} we have:
\begin{equation}
\delta v(r_d) \cdot r_d \sim 
\left(\frac{r_d}{R} \right)^h \delta_R v \cdot r_d \sim \nu.
\end{equation}
Inserting the last expression in the definition of $D_{1,q}(R)$, we
finally have:
\begin{eqnarray}
D_{1,q}(R)& \sim& \int d\mu(h) \left(\delta_R v \right)^{q+2}
R^{-2} \cdot
\nonumber \\
& & \cdot \left( \frac{\nu}{R \delta_R v} 
\right)^{\frac{2 \left( h-1\right)+3-D \left(h \right)}{1+h}} 
\sim  \frac{S_{q+3}(R)}{\nu R}
\label{pippo}
\end{eqnarray}
where we have used the fact that the multifractal
process is such that $\nu \langle (\partial_x v)^2
 \rangle \rightarrow O(1)$
in the limit $\nu \rightarrow 0$. Expression
(\ref{pippo}) coincides with 
 the prediction given in \cite{pro2}.
 The above
computation are easily
generalized for any  
$ \langle  (\partial_xv)^{p} (\delta_R v)^q  \rangle$.

Finally, let us remark that the standard multiplicative
process may not be the end of the story, 
i.e. the dynamics  may be more complex
than what here summarized.\\
 For example, one cannot exclude that
also  sub-leading (with respect to the 
multiplicative ansatz) dynamical processes are acting in the energy
transfer from large to small scales. This dynamical corrections
 must be either negligible with respect to the geometrical
constraints or, at the best, of the same order.

A possible further investigation of such issue would be to 
perform a wavelet analysis of experimental turbulent data.
>From this analysis one may hope to minimize geometrical
constraints focusing only on the dynamical transfer properties.

Other possible candidates to investigate the above problem are
shell models for turbulence, where
geometrical constraints do not affect the energy cascade mechanism.
Work in both directions is in progress.

We acknowledge useful discussions with A.L. Fairhall, V. L'vov and I. 
Procaccia. M. Pasqui is kindly acknowledged for his help
in the analysis of the  synthetic signal. 
We are indebted to S. Ciliberto, R. Chavarria and
 Y. Gagne for having allowed us 
the access to the experimental data.
L.B. and F.T have been supported by INFM (PRA TURBO).

\centerline{FIGURE CAPTIONS}

\noindent
{\bf FIGURE 1}:\\
 Experimental and numerical 
$ F_{p,q}(r,R)/S_{p+q}(R) \cdot S_p(r)$
at fixed $r$ and changing the  large scale $R$. Circles correspond
to $p=2,q=2$, diamonds to $p=4,q=2$ for the experimental data.
Squares correspond to $p=2,q=2$ and triangles to $p=4,q=2$
for the synthetic signal. 
Small scale $r$ is fixed to $ r=16 $ in units of the Kolmogorov scale.
The data for $p=2,q=4$ have been shifted along the vertical axis
for the sake of  presentation.

\noindent
{\bf  FIGURE 2}:\\ 
Experimental $F_{1,2}(r,R)$ at fixed 
$r= 16\,r_d$ and at varying $R$. 
The integral scale $L \sim 1 \times 10^4 \, r_d$.
Let us remark that the observed 
change of sign in the correlation implies 
the presence of at least two power laws. 
The continuos  line 
is the fit in the 
region $r  <  R  < L$ obtained by 
using only the first two terms in (\ref{f1q}).

\begin{thebibliography}{99}

\bibitem{frisch}U. Frisch, {\it Turbulence. The legacy of A.N. Kolmogorov} 
(Cambridge University Press, Cambridge, 1995). 
\bibitem{pro1} V.S. L'vov and I. Procaccia; 
{\it Phys. Rev. Lett} {\bf 76} 2896 {1996}.

\bibitem{pro2} V.S. L'vov and I. Procaccia; 
{\it Phys. Rev. E} {\bf 54} 6268 {1996}.

\bibitem{sreene}A.L. Fairhall, B. Druva, V.S. L'vov, 
I. Procaccia and K.S. Sreenivasan, {\it Phys. Rev. Lett}
(1997) to appear.
\bibitem{physicad96}R. Benzi, L. Biferale, S.Ciliberto, M.V. Struglia
and R.Tripiccione; {\it Physica D} {\bf 96} 162 (1996).
\bibitem{bbcpvv} R. Benzi, L. Biferale, A. Crisanti, G. Paladin,
M. Vergassola and A. Vulpiani; {\it Physica D} {\bf 65} 352
(1993).
\bibitem{bbw} R. Benzi, L. Biferale and A. Wirth; {\it
               Phys. Rev. Lett. } {\bf 78} 4926 (1997).
\bibitem{fv} U. Frisch and M. Vergassola; {\it Europhys. Lett..}
{\bf 14} 439 (1991).
\end{thebibliography}
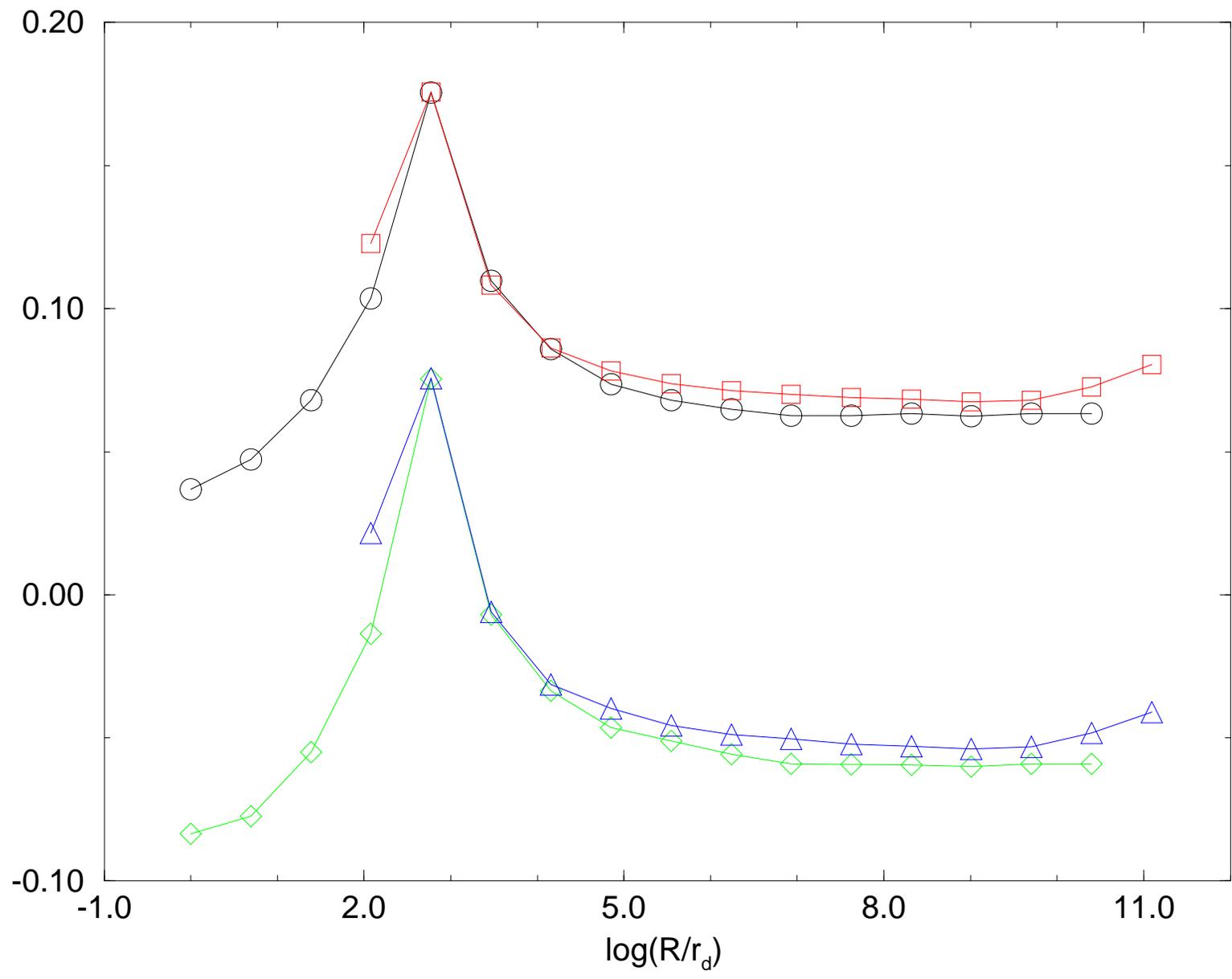
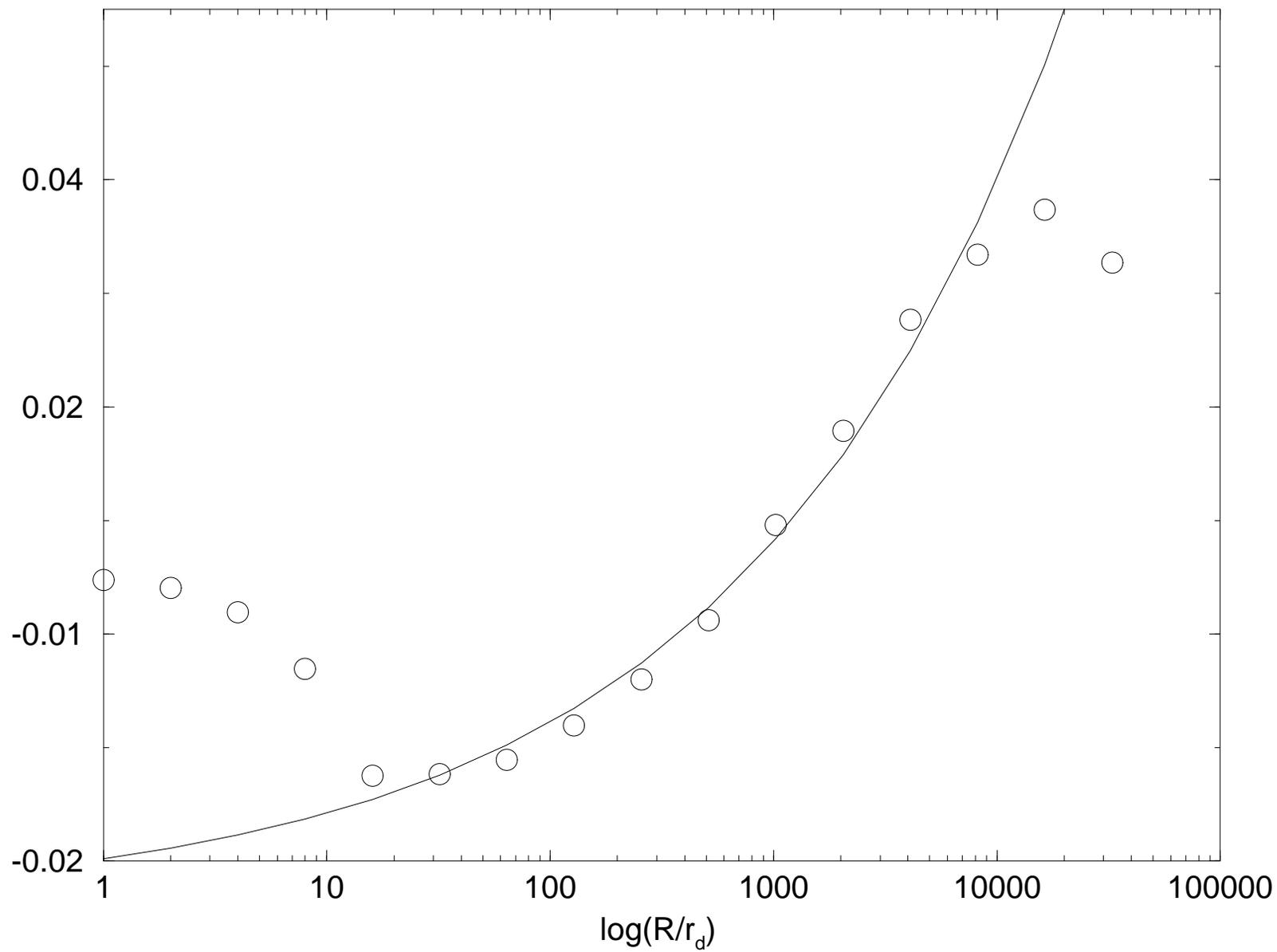
\end{document}